\pdfoutput=1

\documentclass[aps,showpacs,preprintnumbers,amsmath,amssymb,
eqsecnum, twocolumn, tightenlines
]{revtex4}

\usepackage{graphicx}

\newcommand{\be}{\begin{eqnarray}}
\newcommand{\ee}{\end{eqnarray}}

 \newcommand{\gsim}{\mathrel{\hbox{\rlap{\lower.55ex \hbox {$\sim$}}
                   \kern-.3em \raise.4ex \hbox{$>$}}}}
\newcommand{\lsim}{\mathrel{\hbox{\rlap{\lower.55ex \hbox {$\sim$}}
                   \kern-.3em \raise.4ex \hbox{$<$}}}}

\newcommand{\ba}{\begin{eqnarray}}
\newcommand{\ea}{\end{eqnarray}}

\begin{document}

%\twocolumn[\hsize\textwidth\columnwidth\hsize\csname @twocolumnfalse\endcsname

\title{ Monitoring parton equilibration in heavy ion collisions via dilepton polarization}

\author { Edward Shuryak}
\address { Department of Physics and Astronomy, State University of New York,
Stony Brook, NY 11794}
\date{\today}

\begin{abstract}
In this note we discuss how angular distribution of the dileptons produced in heavy ion collisions at RHIC/LHC energies can
provide an information about a degree of local equilibration of the quark-gluon plasma produced
at different invariant mass regions.
\end{abstract}
\maketitle

\section{Introduction}
  The issue of parton equilibration in heavy ion collisions is an area of very active research. 
  As it is well known, successful hydrodynamical description of the elliptic flows 
  \cite{Teaney:2000cw,Teaney:2001av,Huovinen:2001cy,Hirano:2002ds}
  implies that the beginning of (transverse) hydrodynamical expansion cannot start later than $\sim 1/2 \, fm/c$ after the collision moment. Perturbative mechanisms such as e.g. ``bottom-up" equilibration discussed in \cite{Baier:2000sb} have
  difficulties explaining how can it happen so rapidly. On the other hand, applications of the AdS/CFT
 language \cite{Lin:2008rw,Chesler:2008hg,Chesler:2010bi,janik2} naturally ascribe the thermalization time to the ``infall time" into an emerging black hole horizon, which is of the order of
its position in the holographic coordinate  $\sim 1/\pi T_i\sim 0.2 fm/c$. 
  
  Recent studies \cite{janik2,Chesler:2010bi} have followed a set of arbitrarily 
  chosen initial conditions through numerical solution of the Einstein equations. 
  At late time a convergence with a hydrodynamical description is observed, as expected.
   A somewhat surprising    finding is that
  agreement with viscous hydrodynamics is reached when the anisotropy is still quite large.
   We would like therefore to distinguish  the {\em ``hydronization"} \footnote{A term appeared in a discussion at KITP, to my recollection  used by D.Gross first.} time, at which local stress tensor $T^{\mu\nu}$ agrees with hydrodynamical one
  and  the {\em ``anisotropization"} time, at which all distributions become  local (independent on gradients) and thus isotropic. (Both with a prescribed accuracy, of course.)
  
  This letter is not however about theory of equilibration, but about experimental ways to monitor
   it in experiment. Its idea is known in general, but in this short note i would like to provide some
   numerical illustrations of the magnitude of the effect which can be observed in RHIC/LHC 
   heavy ion experiments. 
   
   Let us on the onset remind standard  terminology to be used below. The sources of the dileptons
 are split  into three categories: \\(i)  instantaneous parton annihilation, known as the Drell-Yan process;
 \\   (ii) the pre-equilibrium stage, after the nuclei pass each other; \\
 (iii) equilibrated stage, in which matter is assumed to be local and isotropic.
   
 \section{Angular anisotropy}
It is well known that when spin-1/2 particles (such as  quarks) annihilate and produce
lepton pairs, the cross section is not isotropic but has the following form
\be  {d\sigma \over d\Omega_k}\sim (1+cos^2\theta_k)\ee
where the subscript correspond to a momentum $k$ of,say, the positively charged lepton.
This distribution is, or example, observed in the so called Drell-Yan pairs from stage (i), produced by 
the instantaneous annihilation of the quark-antiquark partons into dileptons. At high energies the partons naturally are collinear to the beams.

\begin{figure}[t]
\begin{center}
\includegraphics [height=6.cm]{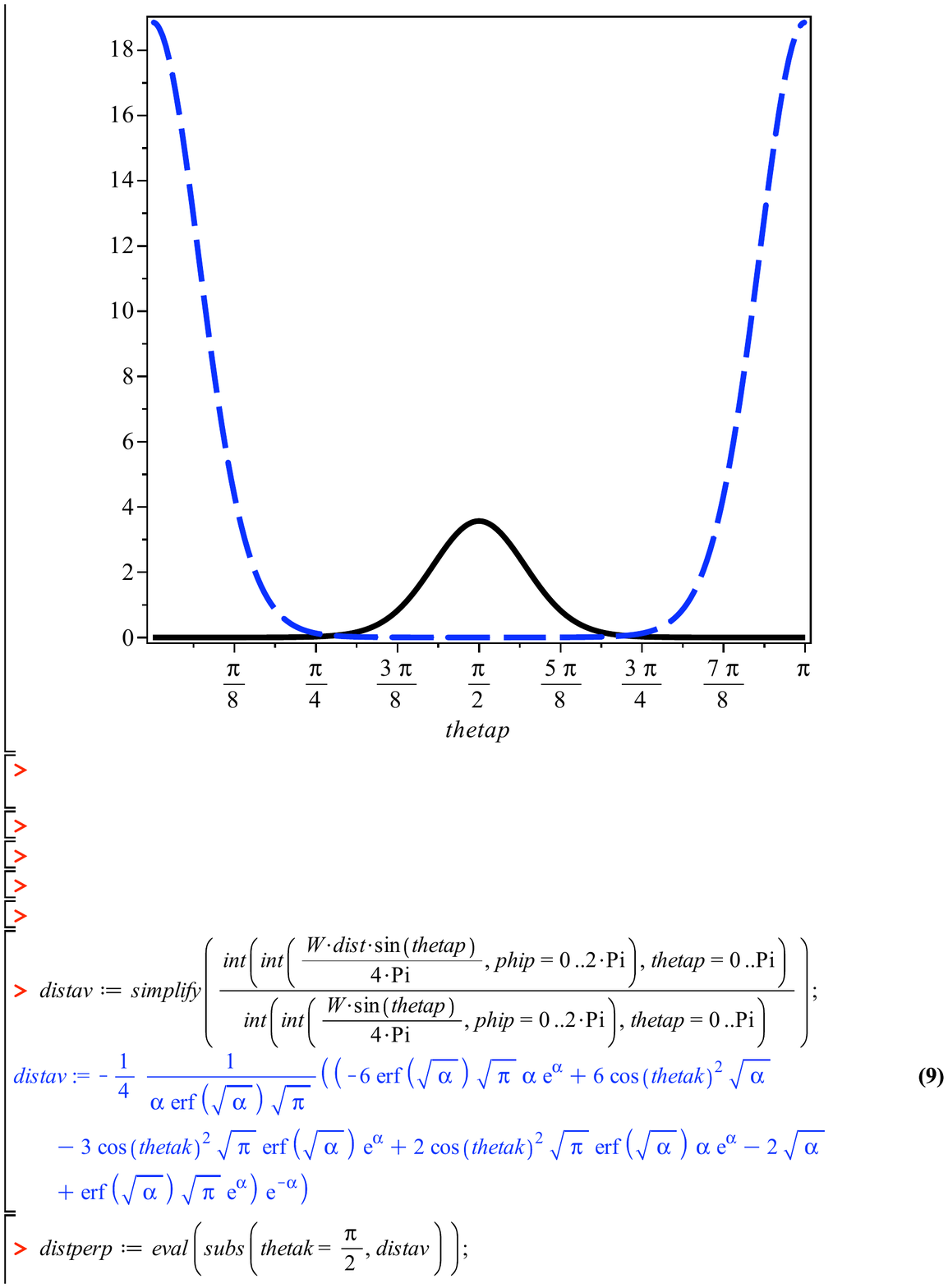}
%\bigskip
\caption{(Color online) Two examples of the angular distribution in polar angle $theta$.  The (black) solid curve corresponds to $\\alpha=10 $,  and the (blue) dashed one to  $\\alpha=-10 $.
}
\label{fig_angular}
\end{center}
\end{figure}

For illustration, let us take a particularly simple one-parameter angular distribution
\be W\sim exp[-\alpha cos^2\theta_p]    \label{W} \ee
with one parameter $\alpha$. 
The subscript $p$ reminds us that this angle is of the colliding partons, not final leptons.
Fig.\ref{fig_angular} shows two opposite examples of (normalized)
distributions.

Let us now calculate the distribution of the dileptons corresponding to he distribution (\ref{W})
\ba 
{d\sigma \over d\Omega_k}&=& { 1 \over 4 erf(\sqrt{\alpha}) \sqrt\pi \alpha } 
 [6 erf(\sqrt{\alpha}) \sqrt\pi \alpha  \nonumber \\
&& +2 \sqrt{\alpha}e^{-\alpha}-erf(\sqrt{\alpha}) \sqrt\pi  \nonumber \\
&& + cos^2\theta_k(-6\sqrt{\alpha}e^{-\alpha} +3\sqrt\pi erf(\sqrt{\alpha}) \nonumber \\
&&-2\sqrt\pi erf(\sqrt{\alpha}) \alpha)]\sim 1+a(\alpha) cos^2\theta_k
\ea
The last expression is a definition of the effective parameter $a(\alpha)$, which we plot in Fig.\ref{fig_angular}.
Note that large negative values of the $\alpha$, corresponding to partons collimated near the beam direction
and Drell-Yan process
$a\approx 1$, as already noticed. 

On the other hand, the second stage of the collision (ii) is characterized by the longitudinal pressure 
smaller than the transverse one. One may understand that because such anisotropic  parton distribution with small differences
in longitudinal momenta is produced by a ``self-sorting" process, in which partons with different rapidities get
spatially separated after the collision. We thus expect at this stage large negative $\alpha$,
in terms of the parameterization we use.  One then finds that the corresponding asymptotic value 
of the anisotropy to be   \be a(\alpha\rightarrow\infty)= -1/3 \ee

\begin{figure}[b]
\begin{center}
\includegraphics [height=6.cm]{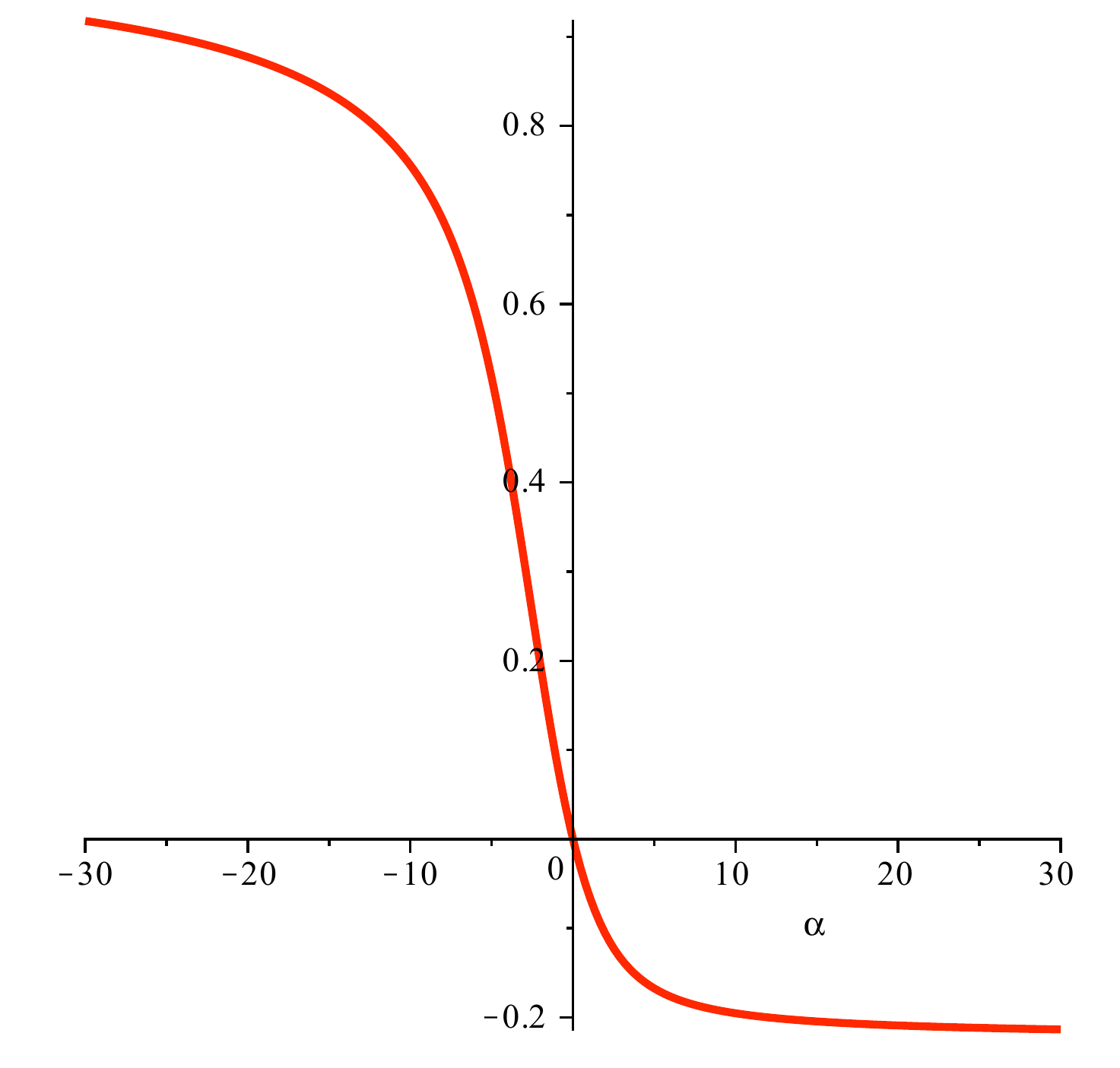}
%\bigskip
\caption{ (Color online) Dependence of the dilepton polarization parameter $a$ on the parton distribution
parameter $\alpha$}
\label{fig_a}
\end{center}
\end{figure}

Finally, when equilibration is over, at stage (iii) the local distributions get isotropic, $\alpha\rightarrow 0$, and the anisotropy
$a(0)=0$.
  \section{Dilepton invariant mass can be used as a clock}
 Qualitatively it is clear that
the most massive dileptons should come from earlier stages. 
The quantitative relation between the invariant mass and the proper time of the production has been studies already in
the first paper suggesting dileptons as a diagnostic tool for QGP \cite{Shuryak:1978ij}.
 It is convenient to introduce the so called
``temperature profile" function
\ba {dN_{e^+e^-}\over dM}&=&\int d^x W(T(x))=\int \int dT d^4x \delta(T-T(x)) W \nonumber \\
&&=\int dT W(T)\Phi(T)\ee
where $\Phi(T)=\int  d^4x \delta(T-T(x))$ is the space-time volume in a fireball in which the temperature is between $T,T+dT$.
Typically the production rate $W\sim exp(-M/T)$ is exponentially decreasing at small $T$, while the profile, which can be
approximately parameterized in a power form $\Phi(T)\sim 1/T^p$ is strongly increasing toward later times and smaller $T$.
The product  has a sharp maximum at the temperature $T^*=M/p$
\be W(T)\Phi(T) \sim  exp[ -{p^3\over 2M^2}(T-T^*)^2]  \ee
Its width gets small if the power $p$ is large, which is in fact the case,  $p\sim 5-7$ or so. The smallness of this width defines  the
accuracy of this clock.

\section{Summary and discussion}

The Drell-Yan
process (i) dominates for large mass dileptons above the charmonium region $M>4\, GeV$: here $\alpha$ is large and negative and the anisotropy
parameter is $a\approx1$.\\
 The preequilibrium phase  (ii) mostly produce the so called Intermediate mass dileptons (IMD), with $m_\phi< M < m_{J\psi}$. 
here we expect large positive $\alpha$ and negative anisotropy  $a\approx -0.2$.\\
 The later well-equilibrated stage (iii) produces smaller mass dileptions $M<1 \,GeV$ which tend to be unpolarized, with
$\alpha\rightarrow 0, a\rightarrow 0$.

The main message of this note is directed toward experimentalists: the dependence of the anisotropy parameter on the dilepton mass $a(M)$ turns out to be of great interest and thus should be  measured.
As we argue, it will change from 1 to a negative value before going to zero. The most negative value measured in experiment
can be compared to our Fig.\ref{fig_a}, from which one can read effective angular distribution parameter $\alpha$
of the quark distribution.

  Finally, let us end with a message to theorists.
Important simplifying assumption made above is that anisotropy of the stress tensor can be directly
translated into the anisotropy of parton distribution in the plasma.  This would be true in 
kinetic (weak coupling) description of QGP, but unfortunately strongly coupled sQGP
is much more complex, in particular it does not allow for partonic quasiparticle description at all.
Furthermore, it was emphasized  \cite{Lin:2008rw} that in out-of-equilibrium setting  the average stress tensor 
(given by a one-point observer) contains information different from what is provided by the non-local (two-point, or spectral) observers. Basically the same message comes out from more recent study \cite{CaronHuot:2011dr}.
It is also not obvious that spectral information in scalar or graviton
correlators are the same as in the vector channel, relevant for dileptons. Thus the latter
should be calculated.

\vskip .25cm {\bf Acknowledgments.} This note is a result of a discussion which took place at the KITP program on novel numerical methods, ADS/CFT etc  in January-March 2012. I am grateful to KITP for the
support  during my stay there.

% bibl from my other paper, to used partly

%\end{narrowtext}


\begin{thebibliography}{99}

%\cite{Teaney:2000cw}
\bibitem{Teaney:2000cw} 
  D.~Teaney, J.~Lauret and E.~V.~Shuryak,
  %``Flow at the SPS and RHIC as a quark gluon plasma signature,''
  Phys.\ Rev.\ Lett.\  {\bf 86}, 4783 (2001)
  [nucl-th/0011058].
  %%CITATION = NUCL-TH/0011058;%%

%\cite{Teaney:2001av}
\bibitem{Teaney:2001av} 
  D.~Teaney, J.~Lauret and E.~V.~Shuryak,
  %``A Hydrodynamic description of heavy ion collisions at the SPS and RHIC,''
  nucl-th/0110037.
  %%CITATION = NUCL-TH/0110037;%%

%\cite{Huovinen:2001cy}
\bibitem{Huovinen:2001cy} 
  P.~Huovinen, P.~F.~Kolb, U.~W.~Heinz, P.~V.~Ruuskanen and S.~A.~Voloshin,
  %``Radial and elliptic flow at RHIC: Further predictions,''
  Phys.\ Lett.\ B {\bf 503}, 58 (2001)
  [hep-ph/0101136].
  %%CITATION = HEP-PH/0101136;%%

%\cite{Hirano:2002ds}
\bibitem{Hirano:2002ds} 
  T.~Hirano and K.~Tsuda,
  %``Collective flow and two pion correlations from a relativistic hydrodynamic model with early chemical freezeout,''
  Phys.\ Rev.\ C {\bf 66}, 054905 (2002)
  [nucl-th/0205043].
  %%CITATION = NUCL-TH/0205043;%%

%\cite{Baier:2000sb}
\bibitem{Baier:2000sb} 
  R.~Baier, A.~H.~Mueller, D.~Schiff and D.~T.~Son,
  %``'Bottom up' thermalization in heavy ion collisions,''
  Phys.\ Lett.\ B {\bf 502}, 51 (2001)
  [hep-ph/0009237].
  %%CITATION = HEP-PH/0009237;%%

%\cite{Lin:2008rw}
\bibitem{Lin:2008rw} 
  S.~Lin and E.~Shuryak,
  %``Toward the AdS/CFT Gravity Dual for High Energy Collisions. 3. Gravitationally Collapsing Shell and Quasiequilibrium,''
  Phys.\ Rev.\ D {\bf 78}, 125018 (2008)
  [arXiv:0808.0910 [hep-th]].
  %%CITATION = ARXIV:0808.0910;%%

%\cite{Chesler:2008hg}
\bibitem{Chesler:2008hg} 
  P.~M.~Chesler and L.~G.~Yaffe,
  %``Horizon formation and far-from-equilibrium isotropization in supersymmetric Yang-Mills plasma,''
  Phys.\ Rev.\ Lett.\  {\bf 102}, 211601 (2009)
  [arXiv:0812.2053 [hep-th]].
  %%CITATION = ARXIV:0812.2053;%%

%\cite{Chesler:2010bi}
\bibitem{Chesler:2010bi} 
  P.~M.~Chesler and L.~G.~Yaffe,
  %``Holography and colliding gravitational shock waves in asymptotically AdS_5 spacetime,''
  Phys.\ Rev.\ Lett.\  {\bf 106}, 021601 (2011)
  [arXiv:1011.3562 [hep-th]].
  %%CITATION = ARXIV:1011.3562;%%

\bibitem{janik2}
  M.~P.~Heller, R.~A.~Janik and P.~Witaszczyk,
  %``The characteristics of thermalization of boost-invariant plasma from
  %holography,''
  arXiv:1103.3452 [hep-th].
  %%CITATION = ARXIV:1103.3452;%%

%\cite{Shuryak:1978ij}
\bibitem{Shuryak:1978ij} 
  E.~V.~Shuryak,
  %``Quark-Gluon Plasma and Hadronic Production of Leptons, Photons and Psions,''
  Phys.\ Lett.\ B {\bf 78}, 150 (1978)
  [Sov.\ J.\ Nucl.\ Phys.\  {\bf 28}, 408 (1978)]
  [Yad.\ Fiz.\  {\bf 28}, 796 (1978)].
  %%CITATION = PHLTA,B78,150;%%
  
  %\cite{CaronHuot:2011dr}
\bibitem{CaronHuot:2011dr} 
  S.~Caron-Huot, P.~M.~Chesler and D.~Teaney,
  %``Fluctuation, dissipation, and thermalization in non-equilibrium AdS_5 black hole geometries,''
  Phys.\ Rev.\ D {\bf 84}, 026012 (2011)
  [arXiv:1102.1073 [hep-th]].
  %%CITATION = ARXIV:1102.1073;%%

\end{thebibliography}
\end{document}